\documentclass[9pt,twocolumn,twoside]{osajnl}

\journal{optica} 

\setboolean{shortarticle}{false}
\usepackage{lipsum}
\usepackage{physics}
\usepackage{siunitx}
\usepackage{float}
\usepackage{blindtext}
\usepackage{graphicx}
\usepackage{color}
\usepackage{units}
\usepackage{dsfont}
\usepackage{hyperref}
\usepackage{svg}
\usepackage{amsmath}
\usepackage{gensymb} 

\title{Multiply-resonant Second-harmonic Generation using Surface Lattice Resonances in Aluminum Metasurfaces}

\author[1]{Timo Stolt}
\author[1]{Anna Vesala}
\author[2]{Heikki Rekola}
\author[2]{Petri Karvinen}
\author[2]{Tommi~K.~Hakala}
\author[1,*]{Mikko J. Huttunen}

\affil[1]{Tampere University, Photonics Laboratory, Physics Unit, Tampere, FI-33014, Finland}
\affil[2]{Faculty of Science and Forestry, Department of Physics and Mathematics, University of Eastern Finland, FI-80101 Joensuu, Finland}

\affil[*]{Corresponding author: mikko.huttunen@tuni.fi}

\begin{abstract}

Nonlinear metamaterials show potential for realizing flat nonlinear optical devices but generally lack in terms of achievable conversion efficiencies. Recent work has focused on enhancing nonlinear processes by utilizing high quality factor resonances, such as collective responses known as surface lattice resonances (SLRs) taking place in periodic metal nanoparticle arrays. Here, we investigate how the dispersive nature of SLRs affects the nonlinear responses of SLR-supporting metasurfaces. Particularly, we measure second-harmonic generation from aluminum nanoparticle arrays and demonstrate that by tilting the sample along two orthogonal directions, the sample can be made multiply-resonant for several pump and second-harmonic signal wavelength combinations. Characterized metasurfaces are estimated to exhibit a  second-order susceptibility value of 0.40 pm/V, demonstrating aluminum as a potential material for nonlinear metasurfaces. 
\end{abstract}

\setboolean{displaycopyright}{true}

\begin{document}

\maketitle

%




\section{Introduction}


Recent developments in miniaturized photonic devices have created a demand for nanoscale nonlinear optical components, which could be potentially addressed by a novel material class known as metamaterials~\cite{Soukoulis2011}. They are artificial structures consisting of subwavelength building blocks, often referred to as meta-atoms. By carefully selecting the meta-atom properties, the bulk metamaterial can exhibit properties not found in natural materials, such as nanoscale phase-engineering capabilities~\cite{Genevet2017, Chen2018a}.
Through these unique properties, metamaterials show potential for realizing novel flat photonic components, such as metalenses and meta-holograms~\cite{Aieta2015,Khorasaninejad2016}. 

In addition to their linear optical properties, also the nonlinear optical properties of metamaterials have been investigated~\cite{Boardman2011,Keren-Zur2018,Krasnok2018,Sain2019}.
For example, plasmonic metamaterials consisting of metal nanoparticles show potential for enabling efficient nonlinear processes in chip-scale devices~\cite{Kauranen2012review,Lee2014}. The optical properties of metal nanoparticles are dominated by the collective oscillations of conduction electrons, known as localized surface plasmons~\cite{Maier2007}.
They exhibit resonant behavior, known as localized surface plasmon resonances (LSPRs), which results in increased local fields near the nanoparticle surface. This local-field enhancement boosts light-matter interaction, including nonlinear processes that scale with high powers of the driving field. Therefore, utilizing LSPRs leads to dramatic enhancements of the nonlinear responses of metal nanoparticles~\cite{Klein2006,Lapine2014, ButetReview2015,Nezami2015, Li2017, Rahimi2018, Huttunen2019review}.
Unfortunately, LSPRs are associated with considerable losses, significantly limiting their potential for nonlinear optics.
Fortunately, by arranging metal nanoparticles into periodic lattices, collective responses known as surface lattice resonances (SLRs) emerge, 
which are associated with much narrower resonance linewidths and higher quality factors ($Q$-factors) than LSPRs~\cite{Wang2018,Kravets2018,Utyushev2021,Saad2021}. This indicates that SLRs are also associated with remarkably stronger local-field enhancements and smaller losses. Consequently, SLRs can be utilized to enhance the nonlinear responses of metasurfaces~\cite{Michaeli2017,Hooper2018,Han2020}.

Most studies on nonlinear plasmonics have focused on singly-resonant metamaterials exhibiting a resonance either at the pump or signal wavelengths of the studied nonlinear processes~\cite{Michaeli2017,Hooper2018,Shen2020}. However, nonlinear processes scale with the local fields at all interacting wavelengths~\cite{BoydBook2020,Maier2007}. Thus, for example a process of second-harmonic generation (SHG) can be enhanced by utilizing multiply-resonant materials, where the resonance enhancement occurs both at the signal and pump wavelengths~\cite{Celebrano2015,Soun2021}.
Furthermore, recent numerical work suggests that multiply-resonant operation based on SLRs could dramatically increase nonlinear responses of plasmonic metasurfaces~\cite{Huttunen2019}.

Here, we experimentally demonstrate multiply-resonant enhancement of SHG in SLR-exhibiting metasurfaces consisting of V-shaped aluminum (Al) nanoparticles. We achieve multiply-resonant operation by tilting the investigated metasurfaces and by utilizing the dispersion of SLRs. The multiply-resonant conditions are fulfilled at several different wavelengths, demonstrating the tunability of SLR-enhanced responses. The measured SH signals correspond to nonlinear susceptibility tensor values of 0.39 pm/V, which is of the same order-of-magnitude as the typical values for traditional nonlinear materials~\cite{BoydBook2020}.

\section{Theory}

\subsection{Surface Lattice Resonances}
The optical properties of metals are governed by the collective oscillations of conduction electrons known as plasmons~\cite{Maier2007}.
At resonant conditions, the strength of light--matter interaction increases, resulting in dramatic changes of, e.g., reflectivity and absorbance of the bulk metal. In the case of metal nanoparticles, plasmons are restricted to the particle surface~\cite{Yao2014}. Therefore, in resonant conditions, incident light is coupled to the local plasmon modes resulting in increased local fields near the nanoparticle surface. This phenomenon is known as LSPR, and it is widely used in many applications of plasmonic metamaterials~\cite{Petryayeva2011,Gu2011,Gao2019}. 
LSPRs are associated with relatively broad linewidths and therefore low $Q$-factors ($Q\,{\sim}\,10$), which indicate extremely short resonance lifetimes but, on the other hand, considerable losses and relatively low local-field enhancements. The low field enhancement can be compensated by using optically dense plasmonic structures and intense pulsed laser sources. Unfortunately, the subsequent strong absorption decrease damage thresholds of the plasmonic structures, significantly limiting the usable input power and thus the strength of the nonlinear responses of plasmonic metamaterials.

\begin{figure}
    \centering
    \includegraphics{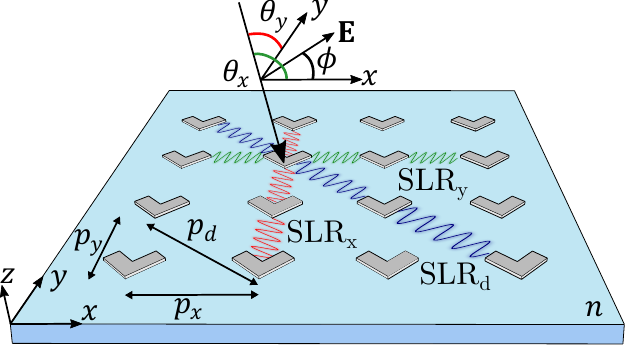}
    \caption{Surface lattice resonance (SLR) modes propagate along the metasurface. Their spectral location depends on numerous factors: Incidence angle ($\theta$), azimuthal angle ($\phi$), refractive index ($n$), and lattice constants $p_x$, $p_y$, and $p_d$. For $y$-polarized light ($\phi=90\degree$), parallel SLRs (green waves) depend on $\theta_x$ and $p_x$,  and for $x$-polarized light ($\phi=0\degree$), the emerging SLRs depend on and $\theta_y$ and $p_y$ (red waves). The diagonal SLRs (blue waves) occur for both polarizations.}
    \label{fig:SLR_schematic}
\end{figure}

A viable approach to decrease losses and increase the interaction strength in plasmonic metamaterials is to utilize SLRs. They are propagating surface modes resulting from radiative coupling of localized surface plasmons in a periodic grating of nanoparticles. They are associated with remarkably high $Q$-factors ($Q\,{\sim}\,1000$)~\cite{Saad2021}, indicating significantly higher local-field enhancements than the ones associated with LSPRs.
Because SLRs result from diffractive properties of the metasurface, their spectral locations are related to the Rayleigh anomalies (RAs) as given by~\cite{MaB2015}:
\begin{equation}
    \lambda_{l,m}(\theta,\phi)=-A_{l,m}(\theta,\phi)+\sqrt{A_{l,m}^2(\theta,\phi)-B_{l,m}(\theta,\phi)} \,\,,
    \label{eq:SLR_full}
\end{equation}
where $l$ and $m$ are the diffraction orders along the Cartesian coordinates of the grating, $\theta$ is the incidence angle in the incidence plane, and $\phi$ is the azimuthal angle that defines the orientation of the incidence plane with respect to the Cartesian coordinates.
For a rectangular metasurface with lattice constants $p_x$ and $p_y$ along the surface Cartesian coordinates (see Fig.~\ref{fig:SLR_schematic}), the variables $A_{l,m}(\theta,\phi)$ and $B_{l,m}(\theta,\phi)$ are given by 
\begin{align}
        A_{l,m}(\theta,\phi) &=\frac{\sin \theta}{(l/p_x)^2+(m/p_y)^2}\left(\frac{l\sin\phi}{p_x}+\frac{m\cos\phi}{p_y}\right) \label{eq:SLR_A}\,,\\
        B_{l,m}(\theta,\phi) &=\frac{\sin^2\theta-n^2}{(l/p_x)^2+(m/p_y)^2}\label{eq:SLR_B}\,,
\end{align}
where  $n$ is the refractive index of the surrounding material. 

In this work, we focus on in-plane SLRs that occur when the polarization of the incident light and the induced dipoles in the nanoparticles are parallel to the metasurface, i.e., when the incident light is TE polarized~\cite{Volk2019}. These SLR modes propagate along the metasurface in the directions that are not parallel with incident polarization. The most obvious option for nanoparticles to couple is along the direction that is perpendicular to the incident polarization, resulting in surface modes we name as parallel SLR.

Here, we are interested in first-order parallel SLRs for light polarized along either $x$- or $y$-axis of the rectangular metasurface. For $x$-polarized light ($\phi=0\degree$), SLR wavelength depends on $p_y$ and the incidence angle $\theta_y$ on the $yz$-plane (see Fig. \ref{fig:SLR_schematic}) as given by
\begin{equation}
    \lambda_{x}(\theta_y)=\,\lambda_{0,\pm1}(\theta_y)=p_y\left(n \mp \sin\theta_y\right)\,.
    \label{eq:SLR_x}
\end{equation}
For $y$-polarized light ($\phi=90\degree$), a similar condition is found to be
\begin{equation}
    \lambda_{y}(\theta_x)=\,\lambda_{\pm 1,0}(\theta_x)=p_x\left(n \mp \sin\theta_x\right)\,,
    \label{eq:SLR_y}
\end{equation}
where $\theta_x$ is the incidence angle along the $xz$-plane with respect to the metasurface normal.

Plasmon modes in a rectangular lattice can couple also along the diagonal of the lattice unit cell, resulting in diagonal SLRs (blue waves in Fig.~\ref{fig:SLR_schematic}).
At normal incidence ($\theta_x=\theta_y=0\degree$), the diagonal SLRs occur at the same wavelength for both incident polarizations, which for first-order SLRs is given by 
\begin{equation}
    \lambda_{d}=n\frac{p_x p_y}{p_d},
    \label{eq:SLR_diag}
\end{equation}
where $p_d=\sqrt{p_x^2+p_y^2}$ is the diagonal of the metasurface unit cell.

Overall, the spectral location of the SLR depends on the polarization of the interacting wave, the lattice constants $p_x$ and $p_y$, the refractive index of the surrounding material $n$, and the incidence angles $\theta_x$ and $\theta_y$, providing us multiple parameters to control the occurrence of SLRs.

\subsection{Second-harmonic Generation in Multiply-resonant Structures}

The SH responses of metamaterials depend on the local fields $E_{\mathrm{loc}}(\omega)$ and $E_{\mathrm{loc}}(2\omega)$, oscillating at the fundamental and SH frequency, respectively~\cite{BoydBook2020,Maier2007}. 
Therefore, we can write for far-field SH emission $E_{\mathrm{nl}}(2\omega)$ that
\begin{equation}
    E_{\mathrm{nl}}(2\omega)\propto\chi^{(2)}E_{\mathrm{loc}}(2\omega)E_{\mathrm{loc}}^2(\omega),
    \label{eq:E_nlo_1}
\end{equation}
where $\chi^{(2)}$ is the effective nonlinear susceptibility of the metasurface.

The local electric fields in \eqref{eq:E_nlo_1} can be enhanced by utilizing metasurface responses, such as LSPRs and SLRs, at the interacting wavelengths. Most works have considered singly-resonant metasurfaces that exhibit resonances at either signal or more commonly at the fundamental wavelength~\cite{Czaplicki2018,Michaeli2017}. However, the nonlinear responses can be further enhanced by utilizing multiply-resonant metasurfaces that exhibit resonances at both interacting wavelengths. For example, Celebrano et al.~designed metasurfaces consisting of gold nanoparticles exhibiting LSPRs to enhance SHG~\cite{Celebrano2015}.
In this work, we extend the approach to metasurfaces based on SLRs with clear benefits. As mentioned before, SLRs are very dependent on the polarization state and propagation direction of the interacting laser fields. Therefore, we design our metasurfaces for type-I SHG, where both fundamental fields have the same polarization state $k$ while the emitted signal field is associated with a different polarization state $j$. For this process, we can rewrite the \eqref{eq:E_nlo_1} as follows:
\begin{equation}
    E_{nl}(\vb{k}_2,2\omega)\propto\chi^{(2)}_{jkk} E_{\mathrm{loc},j}(\vb{k}_2,2\omega)E_{\mathrm{loc},k}^2(\vb{k}_1,\omega),
\end{equation}
where $\vb{k}_1$ and $\vb{k}_2$ are the wavevectors of fundamental and SH beams, respectively, and  $\chi^{(2)}_{jkk}$ is the corresponding effective nonlinear susceptibility tensor component.

\section{Methods}
\label{sec:methods}
\subsection{Sample Fabrication}

For this work, we fabricated Al nanoparticle arrays with a total area of $300\times300$ \si{\micro\m}.
 The structures were fabricated on a pre-cleaned microscope slide (Schott Nexterion, D263T glass). A 200~nm layer of PMMA-resist (MicroChem, 950k) was spin-coated on top and baked on a hot plate at 180\degree C for 180~s. A 10~nm layer of Al was evaporated on the resist to act as a conductive layer for electron beam lithography.
 
The patterning was done using a Raith EBPG 5000+ 100~kV electron beam lithography system. After patterning the Al layer was removed using a 1\% sodium hydroxide solution. The resist was then developed using a 1:3 mixture of methyl isobutyl ketone and isopropanol (IPA) for 15~s, followed by a 30~s immersion in IPA. The sample was dried with nitrogen and placed in an electron beam evaporator for depositing 30~nm of Al. Finally, a liftoff process was performed by soaking the sample in acetone overnight and gently washing the surface with more acetone using a syringe. This removes the resist and excess metal on top of it, leaving only the nanoparticles on the glass substrate. The sample was then rinsed with IPA and dried with nitrogen.

Before the measurements, we covered the metasurface with index-matching oil and an anti-reflection (AR) coated coverslip with the AR wavelength band at 1000--1300 nm. This way, the nanoparticles were assured to have a homogeneous surrounding, and we avoided any Fabry--P\'erot resonances resulting from multiple reflections from different interfaces present in the fabricated devices.

\subsection{Experiments}
In this work, we characterized both linear and nonlinear optical properties of our metasurfaces using two different experimental setups. Here, only short descriptions are given while further details are described in Supplemental Material.

In order to characterize the linear optical properties of our samples, especially the properties of the occurring SLRs, we measured transmission spectra of our samples. Here, we used a broadband halogen lamp, linear polarizer, and spectrometers to locate SLRs.
We placed the sample on a goniometer on a rotational stage, which was connected to a 3-axis translational stage. This enabled continuous control over sample position and orientation, especially with respect to angles $\theta_y$ and $\theta_x$.
By measuring the transmission spectra at different angles, we extracted the dispersion relation graphs shown in Figs.~\ref{fig:C3_results}~(a)--(b) and \ref{fig:A3_results} (a)--(b).

In our nonlinear experiments, we used an optical parametric oscillator (1000--1300 nm)  pumped with a titanium sapphire femtosecond laser (800 nm, repetition rate 82 \si{\MHz}, pulse duration 200 fs) as a tunable laser source.
We set our laser power to 75 mW using a combination of linear polarizer, achromatic half-wave plate and a reference diode. The laser beam was then weakly focused on the sample using an achromatic lens ($f= 500$ mm), resulting in estimated beam diameter ($1/\mathrm{e}^2$ of maximum intensity) of 75 \si{\micro\m} and peak intensity of 115 \si{\MW\per\cm^2} at the sample plane. Similar to the transmission experiments, we placed the sample on a stage that allows fine-tuning of sample position and accurate control of $\theta_y$ and $\theta_x$. In this setup, the rotation stage was motorized allowing continuous incidence angle scans. The generated SH signal was collected with a photo-multiplier tube. By repeating angle scans for different pump wavelengths, we acquired the ($\lambda,\theta$)-graphs for the SHG emission from studied metasurfaces, shown in Figs.~\ref{fig:C3_results}~(c) and \ref{fig:A3_results}~(c).
\section{Design and Results}

\begin{figure}[ht]
    \centering
    \includegraphics{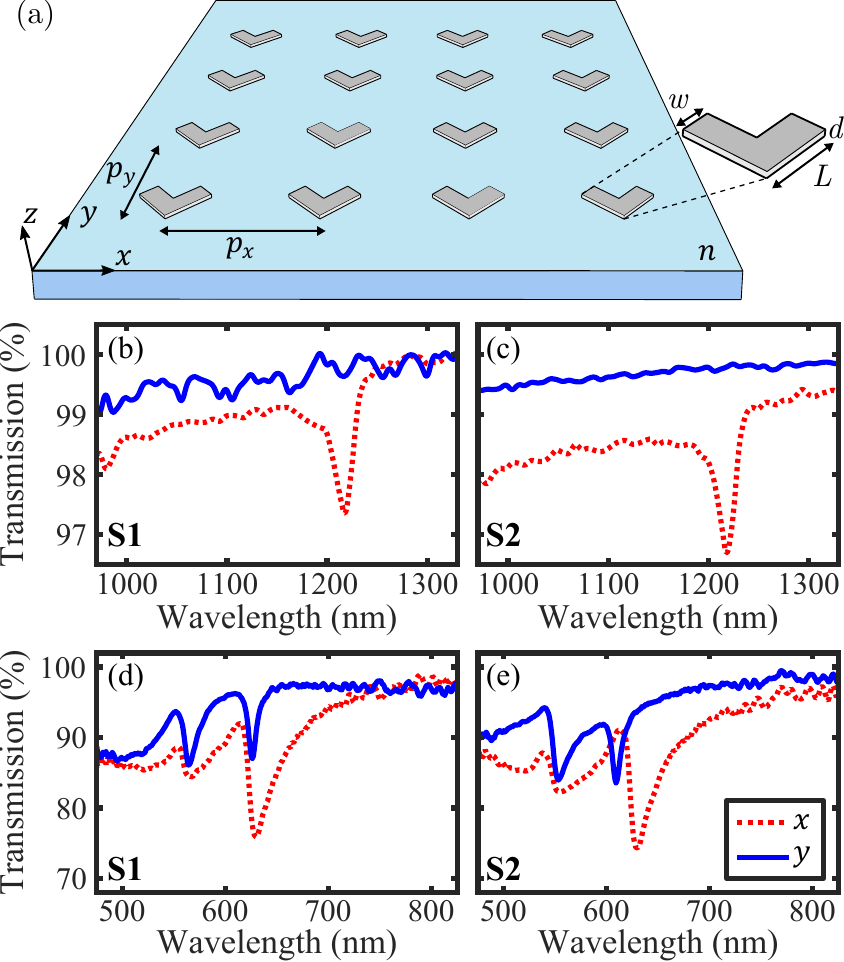}
    \caption{(a) The schematic of the studied metasurfaces, consisting of V-shaped Al nanoparticles fabricated on a glass substrate ($n = 1.51$). Here,  $w= 70$ nm, $L= 100$ nm, and $d= 30$ nm are the arm width, arm length, and the thickness of the Al nanoparticles. (b)--(c) The two studied samples, S1 and S2, had $p_y= 813$ nm, resulting in first-order SLRs at 1220 nm for $x$-polarized light (dotted red lines). (d)--(e) S1 (S2) has $p_x = $ 410 nm (398 nm), resulting in first-order SLR at 626 nm (609 nm) for $y$-polarized light (blue solid lines). Furthermore, S1 (S2) exhibit diagonal SLRs at 560 nm (546 nm).}
    \label{fig:sample_figure}
\end{figure}

Our metasurfaces consisted of V-shaped Al nanoparticles with arm length $L = 100$ nm, arm width $w = 70$ nm, and thickness $d =$ 30 nm (see Fig. \ref{fig:sample_figure}a).
The nanoparticles were fabricated on a glass substrate (refractive index $n=$ 1.51) [See Section \ref{sec:methods} for details]. The nanoparticles were arranged in rectangular lattices with lattice constants $p_y$ and $p_x$ along and orthogonal to the nanoparticle symmetry axis ($y$-axis), respectively (see Fig.~\ref{fig:SLR_schematic}).

Here, we studied two metasurfaces composed of identical nanoparticles but varying in their array periodicities. For sample S1, the periodicities were $p_x=410$ nm and $p_y=813 $, and for sample S2 $p_x=398$ nm and $p_y=813$~nm.
Since both samples are composed of identical nanoparticles, they both exhibit LSPRs centered at 475 nm and 550 nm for $y$- and $x$-polarized light, respectively (see Supplemental Material for more information). The samples also have the same $p_y=813$ nm, and therefore, at normal incidence, they exhibit the first-order SLR for $x$-polarized light at $\lambda_{x}(0\degree)=1220$ nm (see Fig. \ref{fig:sample_figure}b).
The two samples differ in $p_x$, and therefore, also in location of parallel SLRs for $y$-polarized light ($\lambda_y(\theta_x)$) and diagonal SLRs for both polarizations ($\lambda_d$). At normal incidence, $\lambda_y(0\degree)=626$ nm (609 nm) and $\lambda_d=560$ nm (546 nm) for the sample S1 (S2).



Our metasurfaces were designed for the multiply-resonant enhancement of SHG corresponding to $\chi^{(2)}_{yxx}$, i.e., to process with $x$-polarized pump and $y$-polarized SH signal. The multiply-resonant operation is therefore enabled for SHG processes where the pump is coupled to the $x$-polarized SLR and the signal to the $y$-polarized SLR, either parallel or diagonal. Thus, we can write the condition for multiply-resonant operation with resonance wavelengths $\lambda_x(\theta_y)$ and $\lambda_y(\theta_x)$ to be
\begin{equation}
    \lambda_{x}(\theta_y)=2\lambda_{y}(\theta_x)\,,
    \label{eq:multi_res_condition}
\end{equation}
where $\theta_y$ and $\theta_x$ emphasize the fact that we modify the SLR wavelengths by rotating the sample accordingly. This way, the experiment corresponds to a situation where the sample is illuminated at an angle $\theta_y$ and the SH signal is collected at an angle $\theta_x$.

\begin{figure*}[htb]
    \centering
    \includegraphics{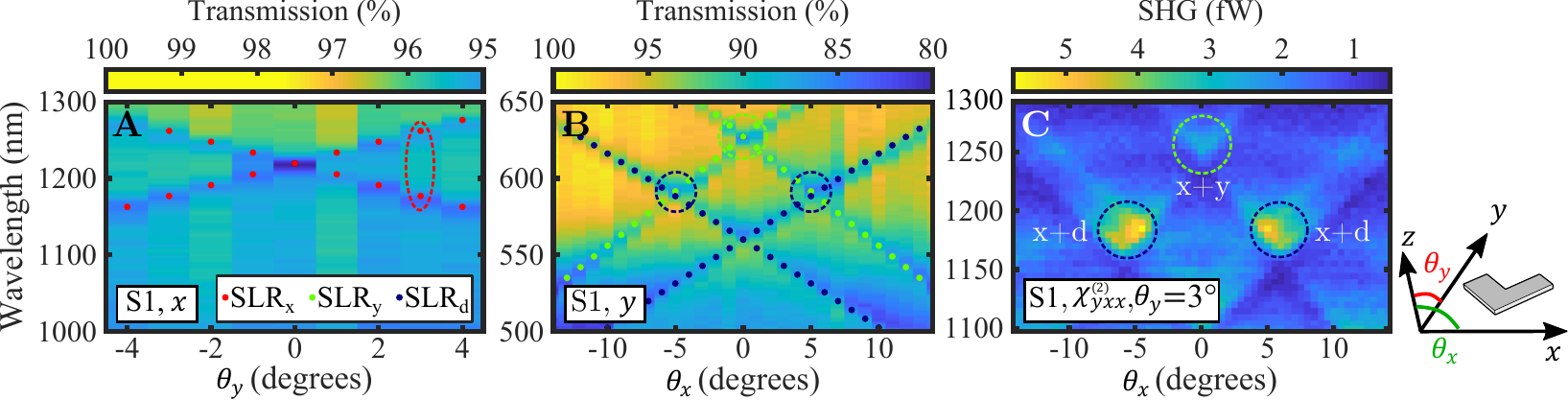}
    \caption{(a)--(b) At normal incidence ($\theta_y=\theta_x=0\degree$), sample S1 exhibits first-order parallel SLRs at 1220 nm and 626 nm for $x$-polarized (red dots) and $y$-polarized (green dots) light, respectively. Additionally, diagonal SLRs (blue dots) occur at 560 nm for normal incidence illumination. By tilting the sample, the SLRs shift from their normal-incidence values. (c) SLRs occurring near pump and signal wavelengths impact the SHG associated with $\chi^{(2)}_{yxx}.$ By setting $\theta_y=3\degree$ (dashed circle in a) and tilting sample with respect to $x$-axis ($\theta_x$), the SH emission is enhanced with three different wavelength--angle combinations (dashed circles in b). At $\theta_x=0\degree$, the parallel SLRs for $x$- and $y$-polarized light enhance SHG at 1250 nm (green circle). At $\theta_x=\pm5\degree$, parallel and diagonal SLRs for $y$-polarized light overlap at 590 nm, and parallel SLR for $x$-polarization occurs at 1176 nm. Combined, these SLRs enhance SHG near 1180 nm (blue circles).}
    \label{fig:C3_results}
\end{figure*}

For sample S1, the multiply-resonant condition of \eqref{eq:multi_res_condition} is not fulfilled at normal incidence, but can be achieved by utilizing the dispersion of the occurring SLRs (see Figs. \ref{fig:C3_results}a and \ref{fig:C3_results}b). Here, we selected $\theta_y=3\degree$ for which S1 exhibit $x$-polarized SLRs at 1176 nm and 1250 nm (red circle in Fig. \ref{fig:C3_results}a). Then, we utilize the dispersion of $y$-polarized SLRs to fulfill the multiply-resonant condition at three different conditions (dashed circles in Fig. \ref{fig:C3_results}b). For pump wavelength 1176 nm, the multiply-resonant condition is fulfilled at $\theta_x=\pm5\degree$, where parallel  SLRs (green dots) and diagonal SLRs (blue dots) overlap at 590 nm. For pump wavelength 1250 nm, the multiply-resonant condition is fulfilled conveniently at $\theta_x=0\degree$, where the fundamental $y$-polarized SLR occurs at 626 nm. These locations of multiply-resonant operation are marked in Figs. \ref{fig:C3_results}a and \ref{fig:C3_results}b with dashed circles.

To demonstrate multiply-resonant operation using the angle--wavelength combinations mentioned above, we set $\theta_y=3\degree$ and measured the SHG ($\theta_x,\lambda$) -spectrum corresponding to $\chi^{(2)}_{yxx}$ (see Fig. \ref{fig:C3_results}c). The SH emission pattern follows the dispersion of $y$-polarized SLRs, and the signal reaches its maximum when multiple SLRs occur at interacting wavelengths (green and blue circles). At these locations, the SH emission is 8-fold, when compared against off-resonance signal. The maximum emission power is 5.7 fW, which corresponds to conversion efficiency of $7.6\times10^{-14}$. By using the method presented in \cite{Herman95,Alloatti2015}, we estimate a value $\chi^{(2)}_{yxx}=0.36 $ \si{pm\per\V} for the sample S1 (see Supplemental Material for calculation details).

However, the maximum SH signal is achieved, when diagonal and parallel SLRs overlap at 590 nm. This overlap results in stronger resonance than the separate SLRs. Additionally, the two SLRs near the pump wavelengths 1176 nm and 1250 nm are considerably weak, when compared against, e.g., the normal-incidence SLR at 1220 nm. We cannot therefore confirm undoubtedly that the strong signals observed with the pump wavelength 1176 nm results from multiply-resonant operation or simply from overlapping SLRs at the signal wavelength. To undoubtedly demonstrate multiply-resonant enhancement, we measured the SH response from the sample S2.

\begin{figure*}[htb]
    \centering
    \includegraphics{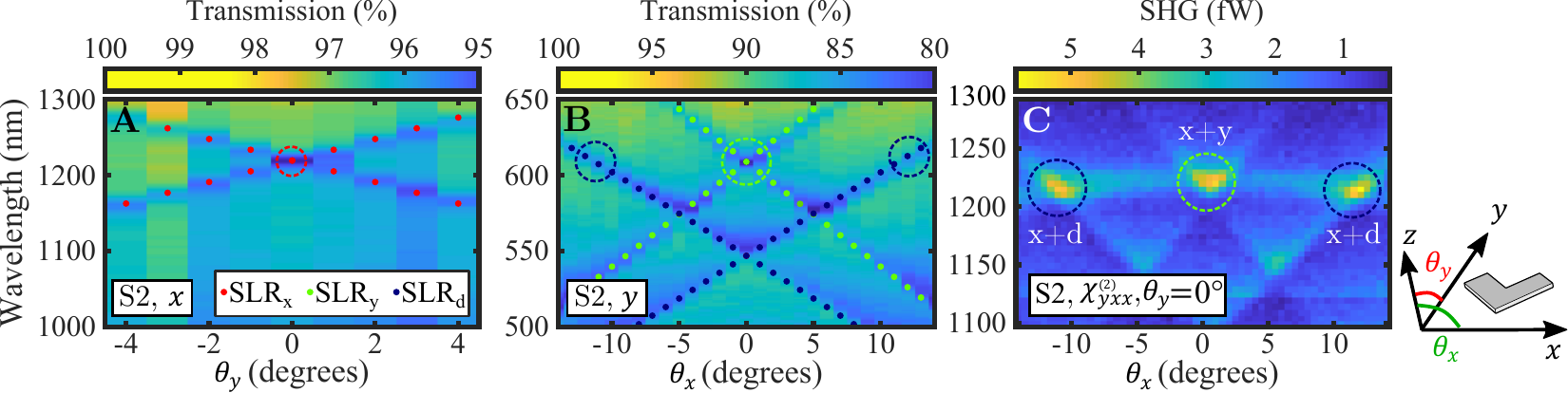}
    \caption{(a)--(b) At normal incidence, sample S2 exhibits first-order parallel SLRs at 1220 nm and 609 nm for $x$-polarized (red dots) and $y$-polarized (green dots) light, respectively. Sample exhibits also diagonal SLRs (blue dots) at 546 nm.  Tilting sample along $y$-axis ($\theta_y$) and $x$-axis ($\theta_x$) will shift $x$- and $y$-polarized SLRs from their normal incidence values. By setting $\theta_y=0\degree$, the multiply-resonant condition for SHG associated with $\chi^{(2)}_{yxx}$ ($\lambda_x=2\lambda_y$) is fulfilled when $\theta_x=[0\degree,\pm11\degree]$ (dashed circles). (c) SLRs enhance the second-harmonic emission and the maximum value is achieved at the multiply-resonant condition.
    With incident angle set as $\theta_y=0\degree$, this occurs at the pump wavelength 1220 nm  with three different emission angles $\theta_x$. First, at $\theta_x=0\degree$ (green circle), the multiply-resonant condition is fulfilled with parallel SLRs. At $\theta_x=\pm11$ (blue circles), S2 exhibit diagonal SLRs at 610 nm, therefore fulfilling the multiply-resonant condition.}
    \label{fig:A3_results}
\end{figure*}

For sample S2, the multiply-resonant condition in \eqref{eq:multi_res_condition} is fulfilled at normal incidence ($\theta_y=0\degree$ and $\theta_x=0\degree$) for the pump wavelength of 1220 nm (red circle in Fig. \ref{fig:A3_results} (a) and green circle in (b)). By rotating the sample along the $x$-axis, i.e., by changing $\theta_x$, the diagonal SLRs (red dots) shift from $\lambda_{d}=546$ nm. At $\theta_x=\pm11\degree$ $\lambda_{d}=610$ nm and the multiply-resonant condition is again fulfilled (blue circles in Fig. \ref{fig:A3_results} b).

To confirm the multiply-resonant operation, we measured the SHG ($\theta_x,\lambda$)-spectrum by setting $\theta_y=0\degree$ and scanning over wavelength range of 1000--1300 nm and angle range ($\theta_x$) from  $-15\degree$ to $15\degree$ (see Fig. \ref{fig:A3_results}c).
The SH emission pattern again follows the dispersion of $y$-polarized SLRs and reaches its maximum at multiply-resonant conditions i.e., at 1220 nm when $\theta_x=[0\degree,\pm11\degree]$ (marked with black and red circles). Now, the multiply-resonant enhancement results in 10-fold enhancement and the maximum emission power of 5.8 fW. The measured signal therefore corresponds to conversion efficiency $7.7\times10^{-14}$ and $\chi^{(2)}_{yxx}=0.40 $ \si{pm\per\V}. 

For sample S2, the impact of multiply-resonant operation is more evident. The SH emission with the pump wavelength 1220 nm is visibly enhanced at all angles $\theta_x$, demonstrating the impact of the SLR at the pump wavelength. More importantly, the signal reaches the maximum level only, when the multiply-resonant condition is fulfilled with parallel and diagonal SLRs occurring at 610 nm, marked in Fig. \ref{fig:A3_results}c with green and blue circles, respectively.

\section{Discussion}
Our results demonstrate two things. First, only few studies have characterized the nonlinear properties of Al nanostructures~\cite{Metzger2015,Metzger2017}. This is due to the fact that Al spontaneously forms oxides (Al\textsubscript{2}O\textsubscript{3}), which impacts the plasmonic properties of the Al nanoparticles~\cite{Langhammer2008}. Therefore, many researchers prefer more stable plasmonic nanostructures, such as gold and silver nanoparticles~\cite{RussierAntoine2007}. However, our work utilizes diffractive properties of plasmonic metasurfaces, which are less sensitive to the changes in LSPRs induced by the oxidation of Al nanoparticles.

Second, our results demonstrate the multiply-resonant enhancement of SHG by utilizing only SLRs. Other works have either utilized only LSPRs or singly-resonant structures that exhibit SLRs near either the pump or signal wavelengths. Unfortunately, our results demonstrate only 10-fold on-and-off-resonance enhancement and SH signal levels of 5 fW, which are significantly lower than the corresponding values acquired with singly-resonant SLR-based structures~\cite{Michaeli2017,Hooper2018}
This difference is mostly due to the fact that our samples exhibit relatively weak (extinction $\leq 5 \%$) and low-quality ($Q\sim60$) SLRs at the pump wavelength. Thus, the SLR-induced field enhancements are also relatively weak.
This is mostly due to the fact the nanoparticles themselves were relatively small when compared against the lattice constant $p_y$, which weakens the interparticle coupling.

Despite the relatively low signal levels, our results demonstrate how SHG can be modified by utilizing the dispersion of SLRs. As is shown in Figures \ref{fig:C3_results} and \ref{fig:A3_results}, SLRs near both pump and signal wavelengths enhance the SH response, which reaches its maximum at the multiply-resonant conditions. By tilting the sample accordingly, we can change the multiply-resonant wavelength, i.e., tune the wavelength of the maximum SH response. Such post-fabrication tunability could prove useful for realizing other nonlinear processes in resonant metasurfaces, such as sum-frequency generation, difference-frequency generation, and third-harmonic generation~\cite{Huttunen2019}. Demonstrating these processes could pave the path towards flat and tunable laser sources with the operation band ranging from the ultraviolet to the terahertz (THz) region of the electromagnetic spectrum.

\section{Conclusion}
In summary, we demonstrate multiply-resonant enhancement of second-harmonic generation from Al metasurfaces. The achieved signal levels correspond to the nonlinear susceptibility tensor component value of 0.40 \si{pm\per\V}, which is the same order-of-magnitude than the susceptibility values of conventional nonlinear optical materials, demonstrating the potential of Al metasurfaces for nonlinear optics. Here, we achieve the multiply-resonant enhancement by utilizing collective responses of periodic metal nanoparticle arrays known as surface lattice resonances. Due to the dispersion of surface lattice resonances, we can control the multiply-resonant enhancement by tilting the sample. As a result, we achieve multiply-resonant enhancement with several different combinations of signal wavelength, incidence angle, and signal emission angle, therefore demonstrating tunable second-harmonic generation. Our methods show promise for realizing other nonlinear processes in plasmonic metasurfaces. Such structures could pave the path towards 
flat and tunable nonlinear devices. 

\bibliography{refs.bib}

\section*{Funding}
We acknowledge the support of the Academy of Finland (Grant No. 308596), the Flagship of Photonics Research and Innovation (PREIN) funded by the Academy of Finland (Grants No. 320165 and 320166). 
TS acknowledges also Jenny and Arttu Wihuri Foundation for doctoral research grant. TKH acknowledges Academy of Finland project number (322002).

\section*{Acknowledgements}
Authors acknowledge Jarno Reuna for providing anti-reflection coatings.

\section*{Disclosures}
The authors declare no conﬂicts of interest.

\newpage
\onecolumn
\newpage
\vspace{2cm}
\begin{center}
\textbf{\Huge Supplemental Material for \\
"Backward phase-matched second-harmonic generation from stacked metasurfaces"}
\end{center}
\setcounter{equation}{0}
\setcounter{figure}{0}
\setcounter{table}{0}
\setcounter{page}{1}
\setcounter{section}{0}
\makeatletter
\renewcommand{\theequation}{S\arabic{equation}}
\renewcommand{\thefigure}{S\arabic{figure}}
\renewcommand{\bibnumfmt}[1]{[#1]}
\renewcommand{\citenumfont}[1]{#1}

\section{Experimental Setups}

\subsection{Linear Setup}
\begin{figure}[h]
    \centering
    \includegraphics{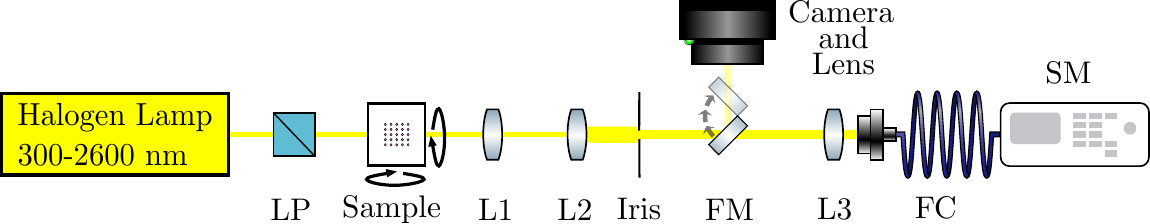}
    \caption{The setup we used in transmission experiments: a broadband halogen lamp, linear polarizer (LP), lenses (L1, $f= 19.5$ \si{\mm}; L2, $f= 75 $ \si{\mm}; L3, $f= 4.3$ \si{\mm}), flip mirror (FM), fiber coupling (FC), and spectrometer (SM).To fine-tune the sample orientation and position, we placed the sample on stage composed of rotational and translation stages, and of a goniometer.}
    \label{fig:setup_linear}
\end{figure}

To characterize the linear optical properties of our samples, we used the setup illustrated in Fig. \ref{fig:setup_linear}. The broadband beam was linearly polarized along the desired axis and aimed at the sample. We placed the sample on a goniometer, which was connected to a stage composed of translational and rotational components. The two achromatic lenses L1 and L2 had two functions. First, they imaged the sample on an iris, to which a camera was focused, allowing us to select the correct area for measuring. Second, they expanded the light beam, which is limited by the iris.  Therefore, only the light propagating through the sample and along the optical axis of the setup was collected. The limited beam was then focused by a lens to an optical fiber that guided the signal to a detector. To study the angle-dependence of SLRs in our metasurfaces, we used a broadband halogen lamp (SLS201 300-2600 nm, Thorlabs) as a light source and AvaSpec-ULS-RS-TEC (Avantes) and  NIR128L-1.7 (Control Development) spectrometers for visible and near-infrared regions, respectively. 

\subsection{Nonlinear Setup}

\begin{figure}
    \centering
    \includegraphics{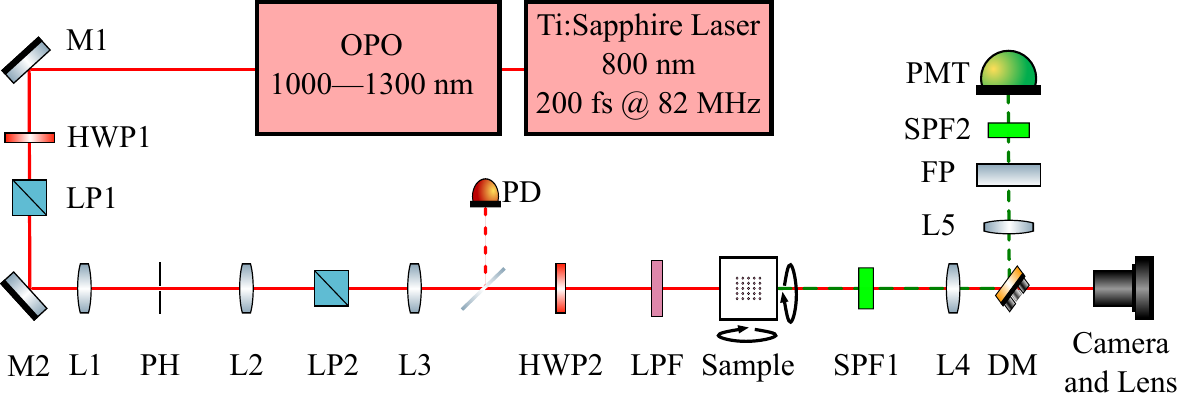}
    \caption{The setup we used in the nonlinear experiments: Ti:Sapphire femtosecond laser, optical parametric oscillator (OPO), mirrors (Ms), linear polarizers (LPs), half-wave plates (HWPs), lenses (L1, $f= 30$ mm; L2, $f= 150$ mm; L3, $f= 500$ mm; L4, $f= 30$ mm; L5, $f= 150$ mm), photo diode (PD), long-pass filter (LPF), short-pass filters (SPFs), dichroic mirror (DM), film polarizer (FP), photo-multiplier tube (PMT), and a camera-lens combination. To fine-tune the sample orientation and position, we placed the sample on a stage composed of rotational and translation stages, and of a goniometer.}
    \label{fig:setup_shg}
\end{figure}

To measure the SH response of our samples, we used the setup illustrated in Fig \ref{fig:setup_shg}. As a tunable laser source, we had an optical parametric oscillator (OPO, 1000\textemdash1300 nm) pumped by a titanium sapphire femtosecond laser (800 nm, repetition rate 82 \si{\MHz}, pulse duration 200 fs). With a combination of two linear polarizers (LPs), a motorized half-wave plate (HWP), and a reference photonic diode (PD), we controlled laser power during the measurements.
A second HWP was used to set the laser polarization along the desired sample axis. Then, the beam was focused on the sample using an achromatic lens. Here, we placed the sample on a similar sample holder that in the linear experiments. Additionally, the sample was placed between a long-pass filter (LPF) and a short-pass filter (SPF) to ensure that only OPO beam was focused on the sample and that only the SH signal generated at the sample was collected.
The SH signal was collected by an achromatic lens and then guided to the detection path with a dichroic mirror (DM).The remaining OPO beam propagated through the DM to a CMOS camera combined with a lens (MVL50M23) that we used to image the sample and select the right metasurface. The signal guided to the detection path was focused by an achromatic lens to photo-multiplier tube (PMT) signal detection. There, another SPF and a film polarizer ensured that only the correct signal wavelengths and polarization would reach the PMT.

\section{Numerical Estimates}
\subsection{Peak Intensity of the Used Femtosecond Laser}

One of the most crucial parameters in nonlinear optical processes is the intensity of the fundamental laser beam $I_{\omega}=P_{\omega}/A_b$, where $P_{\omega}$ and $A_b$ are the power and the area of the laser beam. To evaluate the peak intensity, we first estimated the beam diameter at the sample plane. According to basic Gaussian optics, the beam diameter of a focused Gaussian beam is given by $d\approx\frac{4\lambda f}{\pi d_0}$, where $\lambda$ is the laser wavelength, $f$ is the focal length of  the lens, and $d_0$ is the beam diameter at the focusing lens~\cite{sveltobook}.

In our setup, the beam diameter ($1/\mathrm{e}^2$ of maximum intensity) at the OPO output is 2 mm, which is increased to 1 cm by lenses L1 and L2. By focusing the laser beam ($\lambda=1200$ nm) with the lens L3 ($f= 500$ mm), the beam diameter at the sample plane then becomes 76 \si{\micro\m}. We also estimated the beam size with a CMOS camera by comparing the laser spot to the sample arrays (area $300\times300\, \si{\micro\m}^2$) and got the exact same beam size. Then, we could estimate the peak intensity of a femtosecond laser pulse ($P_{avg}=75 $mW, $\tau_p=200$ fs, $\nu_{rep}=82$ MHz) as 
\begin{equation}
    I_{peak}\approx\frac{P_{avg}}{1.763A_b\tau_p\nu_{rep}}\approx 115\, \si{\MW\per\cm^2},
\end{equation}
where $A_b=\frac{\pi(d/2)^2}{2}$ is the area of a Gaussian laser beam and the factor 1.763 takes into account the $\mathrm{sech}$-shaped temporal profile.

\subsection{Nonlinear Susceptibility Calculations}
In this article, we estimated the nonlinear susceptibility tensor component $\chi^{(2)}_{yxx}$ using the method detailed in~\cite{Herman95} and later applied for metamaterials~\cite{Alloatti2015,Saad2021NL}. The power of SH signal can be estimated using the following formula:
\begin{equation}
    P_{2\omega}=\frac{2}{c_0\epsilon_0A}\frac{[t^{(1)}_{ms}]^4[t^{(2)}_{ms}]^2[t_{sa}^{(2)}]^2}{n_2^2c_2^2}P_{\omega}^2\left(\frac{2\pi L}{\lambda}\right)^2\left(\frac{1}{2}\chi^{(2)}\right)^2\frac{\mathrm{sinc}^2\psi+R_1+R_2}{1+R_3+R_4},
    \label{eq:herman}
\end{equation}
where $\psi=(2\pi L/\lambda)(n_1c_1-n_2c_2)$, $P_{2\omega}$ and $P_{\omega}$ are the SH and fundamental power, respectively, $A$ is the beam area on the sample plane, $L$ is the length of the nonlinear material, $c_0$  is the speed of light in vacuum, $\epsilon_0$ is the vacuum permittivity,and $\lambda$ is the fundamental beam wavelength.

In \eqref{eq:herman}, $t_{ms}^{(1)}$ and $t_{ms}^{(2)}$ are the transmission coefficient of the metasurface at fundamental and SH frequency, respectively. They are connected to the measured transmission values by $T_{\omega}=\left|t^{(1)}\right|^2$ and $T_{2\omega}=\left|t^{(2)}\right|^2$. Since we only consider the magnitude of the SH signal and do not consider signal propagation in the substrate, we can consider the absolute values of $t^{(1)}$ and $t^{(2)}$ and rewrite the related parts in \eqref{eq:herman}: $[t_{ms}^{(1)}]^4=T_{\omega}^2$ and $[t_{ms}^{(2)}]^2=T_{2\omega}$. The third transmission coefficient $t_{sa}^{(2)}$ describes the substrate-air transmission. Our samples were fabricated on glass ($n=1.51$), resulting in $t_{sa}^{(2)}=1.2076$ and consequently that $[t_{sa}^{(2)}]^2=1.4583$.

The $n_1$ and $n_2$ are the refractive indices at fundamental and SH frequency, respectively, and $c_m=\sqrt{1-(1/n_m)^2\sin^2\theta}$, where $\theta$ is the incident angle. For simplicity, we assume small incident angles, and therefore, $c_1\approx c_2\approx 1$. For our metasurfaces, $L=30$ nm and $\lambda\sim 1200$ nm, from which overall follows that $\psi\approx0$ and $\mathrm{sinc}^2\psi\approx1$.  The terms $R_1$, ..., $R_4$ are related to the multiple reflections on medium surfaces, which we neglect for simplicity, further justified by anti-reflection coatings placed on our samples. Finally, we modify the quotient $P_{\omega}/A$ for Gaussian beams to include a factor of $\ln 2\sqrt{2\ln 2/\pi}/(2\nu_{rep}\tau_p)$. Now, we can rewrite \eqref{eq:herman}:
\begin{equation}
    P_{2\omega}^{(x\rightarrow y)}=\frac{\ln2}{4c_0\epsilon_0\nu_{rep}\tau_p}\sqrt{\frac{2\ln2}{\pi}}\frac{T_{\omega}^2T_{2\omega}[t_{sa}^{(2)}]^2}{n_2^2}\frac{P_{\omega}^2}{A}\left(\frac{2\pi L}{\lambda}\right)^2\left(\chi^{(2)}_{yxx}\right)^2 .
    \label{eq:sh_response}
\end{equation}
For sample S2, using the femtosecond laser and metasurface parameters listed above, and by obtaining $T_{\omega}=0.95$, $T_{2\omega}=0.82$, $P_{\omega}= 75$ mW, and $P_{2\omega}=5.8$ fW from the experimental results, we can calculate that $\chi^{(2)}\approx 0.40$ pm/V.

\section{Spectral Location of Localized Surface Plasmon Resonances}
\begin{figure}[h]
    \centering
    \includegraphics{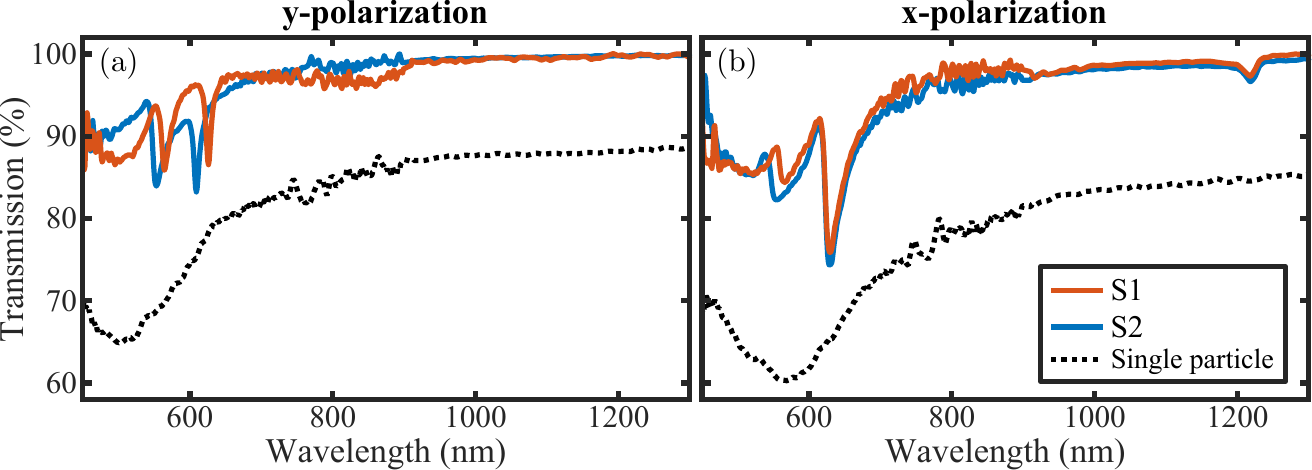}
    \caption{When arranged on a random lattice, the metasurface exhibits only single-particle LSPRs (black dashed lines). For our samples, they occur at 475 nm and 550 nm for (a) $y$- and (b) $x$-polarized light (solid lines), respectively. When the particles are arranged in rectangular lattices, the metasurfaces exhibit SLRs near the Rayleigh Anomaly wavelengths.}
    \label{fig:rand_T}
\end{figure}

To define the spectral locations of single-particle localized surface plasmon resonances (LSPRs), we fabricated a sample with a random lattice configuration with a particle density similar to 300 nm x 300 nm square array. The transmission spectra (dashed black lines in Fig \ref{fig:rand_T}) reveal that LSPRs with linewidths $\sim100$ nm are centered around 475 nm and 550 nm for (a) $y$- and (b) $x$-polarized light, respectively. Comparison against spectra of samples S1 (red solid lines) and S2 (blue solid lines) reveal the impact of LSPRs on surface lattice resonances (SLRs).
First, Rayleigh anomalies (RAs) at longer wavelengths than LSPR result in SLR peaks, e.g., at 1220 nm for $x$-polarized light.
Second, the overlap of RA and LSPR wavelengths results in "dips" in the LSPR peaks at the RA wavelength. For example, sample S2 has second-order parallel RA at 610 nm for $x$-polarized light and a diagonal RA at 536 nm for both polarizations. As these wavelengths overlap with LSPRs, the coupling between diffractive and plasmon modes results in sharp dips in otherwise broad LSPR peaks.

\section{Power-dependence of SHG from Sample S2}
\begin{figure}
    \centering
    \includegraphics{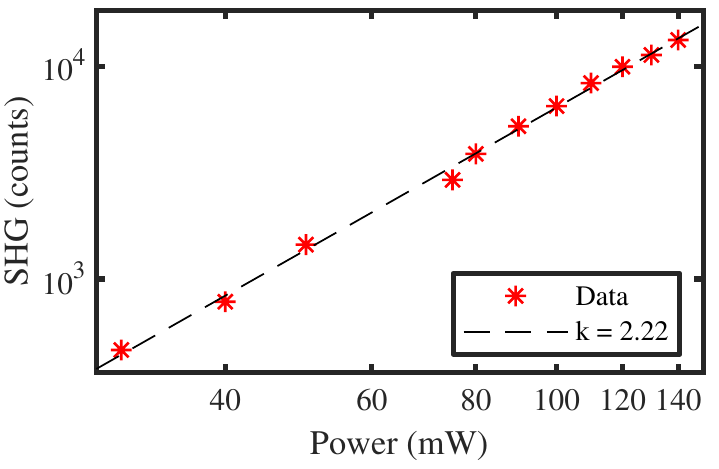}
    \caption{The second-harmonic emission from our samples is square-dependent on the incident power.}
    \label{fig:SHG_vs_P}
\end{figure}

To ensure that the detected signal was indeed SH, we measured the normal-incidence ($\theta_x=\theta_y=0\degree$) SH signal from our sample S2 at 1220 nm using different input powers between 30 and 140 mW. The measured power dependence is plotted in logarithmic scale in Fig \ref{fig:SHG_vs_P}. We fitted a linear function (black dashed line) to the measured data (red stars), which had a slope $k=2.22$ confirming the square-dependence of our signal, i.e., confirming indeed that the signal was SH. Furthermore, and even more importantly, the maximum power was limited by our setup and not by our sample. Therefore, we assume that our sample can withstand high pulse energies. This motivates us to study nonlinear properties with different laser sources such as picosecond lasers that have narrower linewidths but significantly higher pulse energies that would be destructive to plasmonic metasurfaces that utilize LSPRs.

\bibliography{refs}

\begin{thebibliography}{10}
\newcommand{\enquote}[1]{``#1''}

\bibitem{Soukoulis2011}
C.~M. Soukoulis and M.~Wegener, \enquote{{Past achievements and future
  challenges in the development of three-dimensional photonic metamaterials},}
  {\protect\JournalTitle{Nat. Photonics}} \textbf{5}, 523 (2011).

\bibitem{Genevet2017}
P.~Genevet, F.~Capasso, F.~Aieta, M.~Khorasaninejad, and R.~Devlin,
  \enquote{{Recent advances in planar optics: from plasmonic to dielectric
  metasurfaces},} {\protect\JournalTitle{Optica}} \textbf{4}, 139 (2017).

\bibitem{Chen2018a}
S.~Chen, Z.~Li, Y.~Zhang, H.~Cheng, and J.~Tian, \enquote{{Phase Manipulation
  of Electromagnetic Waves with Metasurfaces and Its Applications in
  Nanophotonics},} {\protect\JournalTitle{Adv. Opt. Mater.}} \textbf{6}, 1--25
  (2018).

\bibitem{Aieta2015}
F.~Aieta, M.~A. Kats, P.~Genevet, and F.~Capasso, \enquote{{Achromatic
  metasurfaces by dispersive phase compensation},}
  {\protect\JournalTitle{Science (80-. ).}} \textbf{347}, 1342--1345 (2015).

\bibitem{Khorasaninejad2016}
M.~Khorasaninejad, A.~Ambrosio, P.~Kanhaiya, and F.~Capasso, \enquote{Broadband
  and chiral binary dielectric meta-holograms,} {\protect\JournalTitle{Science
  Advances}} \textbf{2} (2016).

\bibitem{Boardman2011}
A.~D. Boardman, V.~V. Grimalsky, Y.~S. Kivshar, S.~V. Koshevaya, M.~Lapine,
  N.~M. Litchinitser, V.~N. Malnev, M.~Noginov, Y.~G. Rapoport, and V.~M.
  Shalaev, \enquote{{Active and tunable metamaterials},}
  {\protect\JournalTitle{Laser Photonics Rev.}} \textbf{5}, 287--307 (2011).

\bibitem{Keren-Zur2018}
S.~Keren-Zur, L.~Michaeli, H.~Suchowski, and T.~Ellenbogen, \enquote{{Shaping
  light with nonlinear metasurfaces},} {\protect\JournalTitle{Adv. Opt.
  Photonics}} \textbf{10}, 309 (2018).

\bibitem{Krasnok2018}
A.~Krasnok, M.~Tymchenko, and A.~Al{\`{u}}, \enquote{{Nonlinear metasurfaces: a
  paradigm shift in nonlinear optics},} {\protect\JournalTitle{Mater. Today}}
  \textbf{21}, 8--21 (2018).

\bibitem{Sain2019}
B.~Sain, C.~Meier, and T.~Zentgraf, \enquote{{Nonlinear optics in
  all-dielectric nanoantennas and metasurfaces: a review},}
  {\protect\JournalTitle{Adv. Photonics}} \textbf{1}, 1 (2019).

\bibitem{Kauranen2012review}
M.~Kauranen and A.~V. Zayats, \enquote{{Nonlinear plasmonics},}
  {\protect\JournalTitle{Nat. Photonics}} \textbf{6}, 737--748 (2012).

\bibitem{Lee2014}
J.~Lee, M.~Tymchenko, C.~Argyropoulos, P.~Y. Chen, F.~Lu, F.~Demmerle,
  G.~Boehm, M.~C. Amann, A.~Al{\`{u}}, and M.~A. Belkin, \enquote{{Giant
  nonlinear response from plasmonic metasurfaces coupled to intersubband
  transitions},} {\protect\JournalTitle{Nature}} \textbf{511}, 65--69 (2014).

\bibitem{Maier2007}
S.~A. Maier, \emph{{Plasmonics: fundamentals and applications}} (Springer
  Science {\&} Business Media, 2007).

\bibitem{Klein2006}
M.~W. Klein, C.~Enkrich, M.~Wegener, and S.~Linden, \enquote{{Second-harmonic
  generation from magnetic metamaterials},} {\protect\JournalTitle{Science
  (80-. ).}} \textbf{313}, 502--504 (2006).

\bibitem{Lapine2014}
M.~Lapine, I.~V. Shadrivov, and Y.~S. Kivshar, \enquote{{Colloquium: Nonlinear
  metamaterials},} {\protect\JournalTitle{Rev. Mod. Phys.}} \textbf{86},
  1093--1123 (2014).

\bibitem{ButetReview2015}
J.~Butet, P.~F. Brevet, and O.~J. Martin, \enquote{{Optical Second Harmonic
  Generation in Plasmonic Nanostructures: From Fundamental Principles to
  Advanced Applications},} {\protect\JournalTitle{ACS Nano}} \textbf{9},
  10545--10562 (2015).

\bibitem{Nezami2015}
M.~S. Nezami and R.~Gordon, \enquote{{Localized and propagating surface plasmon
  resonances in aperture-based third harmonic generation},}
  {\protect\JournalTitle{Opt. Express}} \textbf{23}, 32006 (2015).

\bibitem{Li2017}
G.~Li, S.~Zhang, and T.~Zentgraf, \enquote{{Nonlinear photonic metasurfaces},}
  {\protect\JournalTitle{Nat. Rev. Mater.}} \textbf{2}, 1--14 (2017).

\bibitem{Rahimi2018}
E.~Rahimi and R.~Gordon, \enquote{{Nonlinear Plasmonic Metasurfaces},}
  {\protect\JournalTitle{Adv. Opt. Mater.}} \textbf{6}, 1--9 (2018).

\bibitem{Huttunen2019review}
M.~J. Huttunen, R.~Czaplicki, and M.~Kauranen, \enquote{{Nonlinear plasmonic
  metasurfaces},} {\protect\JournalTitle{J. Nonlinear Opt. Phys. Mater.}}
  \textbf{28}, 1950001 (2019).

\bibitem{Wang2018}
W.~Wang, M.~Ramezani, A.~I. V{\"{a}}kev{\"{a}}inen, P.~T{\"{o}}rm{\"{a}}, J.~G.
  Rivas, and T.~W. Odom, \enquote{{The rich photonic world of plasmonic
  nanoparticle arrays},} {\protect\JournalTitle{Mater. Today}} \textbf{21},
  303--314 (2018).

\bibitem{Kravets2018}
V.~G. Kravets, A.~V. Kabashin, W.~L. Barnes, and A.~N. Grigorenko,
  \enquote{{Plasmonic Surface Lattice Resonances: A Review of Properties and
  Applications},} {\protect\JournalTitle{Chem. Rev.}} \textbf{118}, 5912--5951
  (2018).

\bibitem{Utyushev2021}
A.~D. Utyushev, V.~I. Zakomirnyi, and I.~L. Rasskazov, \enquote{Collective
  lattice resonances: Plasmonics and beyond,} {\protect\JournalTitle{Reviews in
  Physics}} \textbf{6}, 100051 (2021).

\bibitem{Saad2021}
M.~S. Bin-Alam, O.~Reshef, Y.~Mamchur, M.~Z. Alam, G.~Carlow, J.~Upham, B.~T.
  Sullivan, J.-M. Ménard, M.~J. Huttunen, R.~W. Boyd, and K.~Dolgaleva,
  \enquote{Ultra-high-q resonances in plasmonic metasurfaces,}
  {\protect\JournalTitle{Nature Communications}} \textbf{12}, 974 (2021).

\bibitem{Michaeli2017}
L.~Michaeli, S.~Keren-Zur, O.~Avayu, H.~Suchowski, and T.~Ellenbogen,
  \enquote{Nonlinear surface lattice resonance in plasmonic nanoparticle
  arrays,} {\protect\JournalTitle{Physical Review Letters}} \textbf{118}
  (2017).

\bibitem{Hooper2018}
D.~C. Hooper, C.~Kuppe, D.~Wang, W.~Wang, J.~Guan, T.~W. Odom, and V.~K. Valev,
  \enquote{{Second harmonic spectroscopy of surface lattice resonances},}
  {\protect\JournalTitle{Nano Lett.}} \textbf{19}, 165--172 (2018).

\bibitem{Han2020}
A.~Han, C.~Dineen, V.~E. Babicheva, and J.~V. Moloney, \enquote{Second harmonic
  generation in metasurfaces with multipole resonant coupling,}
  {\protect\JournalTitle{Nanophotonics}} \textbf{9}, 3545--3556 (2020).

\bibitem{Shen2020}
B.~Shen, L.~Liu, Y.~Li, S.~Ren, J.~Yan, R.~Hu, and J.~Qu, \enquote{Nonlinear
  spectral-imaging study of second- and third-harmonic enhancements by
  surface-lattice resonances,} {\protect\JournalTitle{Advanced Optical
  Materials}} \textbf{8} (2020).

\bibitem{BoydBook2020}
R.~W. Boyd, \emph{{Nonlinear optics}} (Academic Press, San Diego, 2020), 4th
  ed.

\bibitem{Celebrano2015}
M.~Celebrano, X.~Wu, M.~Baselli, S.~Gro{\ss}mann, P.~Biagioni, A.~Locatelli,
  C.~{De Angelis}, G.~Cerullo, R.~Osellame, B.~Hecht, L.~Du{\`{o}},
  F.~Ciccacci, and M.~Finazzi, \enquote{{Mode matching in multiresonant
  plasmonic nanoantennas for enhanced second harmonic generation},}
  {\protect\JournalTitle{Nat. Nanotechnol.}} \textbf{10}, 412--417 (2015).

\bibitem{Soun2021}
L.~Soun, B.~Fix, H.~E. Ouazzani, S.~Héron, N.~Bardou, C.~Dupuis, S.~Derelle,
  J.~Jaeck, R.~Haïdar, and P.~Bouchon, \enquote{Experimental demonstration of
  second-harmonic generation in high $\chi^{2}$ metasurfaces,}
  {\protect\JournalTitle{Optics Letters}} \textbf{46}, 1466 (2021).

\bibitem{Huttunen2019}
M.~J. Huttunen, O.~Reshef, T.~Stolt, K.~Dolgaleva, R.~W. Boyd, and M.~Kauranen,
  \enquote{{Efficient nonlinear metasurfaces by using multiresonant high-Q
  plasmonic arrays},} {\protect\JournalTitle{J. Opt. Soc. Am. B}} \textbf{36},
  E30--E35 (2019).

\bibitem{Yao2014}
K.~Yao and Y.~Liu, \enquote{Plasmonic metamaterials,}
  {\protect\JournalTitle{Nanotechnology Reviews}} \textbf{3}, 177--210 (2014).

\bibitem{Petryayeva2011}
E.~Petryayeva and U.~J. Krull, \enquote{Localized surface plasmon resonance:
  Nanostructures, bioassays and biosensing-a review,}
  {\protect\JournalTitle{Analytica Chimica Acta}} \textbf{706}, 8--24 (2011).

\bibitem{Gu2011}
X.~Gu, T.~Qiu, W.~Zhang, and P.~K. Chu, \enquote{Light-emitting diodes enhanced
  by localized surface plasmon resonance,} {\protect\JournalTitle{Nanoscale
  Research Letters}} \textbf{6} (2011).

\bibitem{Gao2019}
P.~F. Gao, Y.~F. Li, and C.~Z. Huang, \enquote{Localized surface plasmon
  resonance scattering imaging and spectroscopy for real-time reaction
  monitoring,} {\protect\JournalTitle{Applied Spectroscopy Reviews}}
  \textbf{54}, 237--249 (2019).

\bibitem{MaB2015}
T.~W. Maß and T.~Taubner, \enquote{Incident angle-tuning of infrared antenna
  array resonances for molecular sensing,} {\protect\JournalTitle{ACS
  Photonics}} \textbf{2}, 1498--1504 (2015).

\bibitem{Volk2019}
K.~Volk, J.~P. Fitzgerald, and M.~Karg, \enquote{In-plane surface lattice and
  higher order resonances in self-assembled plasmonic monolayers: From
  substrate-supported to free-standing thin films,} {\protect\JournalTitle{ACS
  Applied Materials and Interfaces}} \textbf{11}, 16096--16106 (2019).

\bibitem{Czaplicki2018}
R.~Czaplicki, A.~Kiviniemi, M.~J. Huttunen, X.~Zang, T.~Stolt, I.~Vartiainen,
  J.~Butet, M.~Kuittinen, O.~J.~F. Martin, and M.~Kauranen, \enquote{{Less is
  more – enhancement of second-harmonic generation from metasurfaces by
  reduced nanoparticle density},} {\protect\JournalTitle{Nano Lett.}}
  \textbf{18}, 7709--7714 (2018).

\bibitem{Herman95}
W.~N. Herman and L.~M. Hayden, \enquote{Maker fringes revisited:
  second-harmonic generation from birefringent or absorbing materials,}
  {\protect\JournalTitle{J. Opt. Soc. Am. B}} \textbf{12}, 416--427 (1995).

\bibitem{Alloatti2015}
L.~Alloatti, C.~Kieninger, A.~Froelich, M.~Lauermann, T.~Frenzel, K.~Köhnle,
  W.~Freude, J.~Leuthold, M.~Wegener, and C.~Koos, \enquote{Second-order
  nonlinear optical metamaterials: Abc-type nanolaminates,}
  {\protect\JournalTitle{Applied Physics Letters}} \textbf{107} (2015).

\bibitem{Metzger2015}
B.~Metzger, L.~Gui, J.~Fuchs, D.~Floess, M.~Hentschel, and H.~Giessen,
  \enquote{{Strong Enhancement of Second Harmonic Emission by Plasmonic
  Resonances at the Second Harmonic Wavelength},} {\protect\JournalTitle{Nano
  Lett.}} \textbf{15}, 3917--3922 (2015).

\bibitem{Metzger2017}
B.~Metzger, M.~Hentschel, and H.~Giessen, \enquote{{Probing the Near-Field of
  Second-Harmonic Light around Plasmonic Nanoantennas},}
  {\protect\JournalTitle{Nano Lett.}} \textbf{17}, 1931--1937 (2017).

\bibitem{Langhammer2008}
C.~Langhammer, M.~Schwind, B.~Kasemo, and I.~Zorić, \enquote{Localized surface
  plasmon resonances in aluminum nanodisks,} {\protect\JournalTitle{Nano
  Letters}} \textbf{8}, 1461--1471 (2008).

\bibitem{RussierAntoine2007}
E.~Benichou, G.~Bachelier, C.~Jonin, P.~F. Brevet, and C.~B. Lyon,
  \enquote{{Multipolar Contributions of the Second Harmonic Generation from
  Silver and Gold Nanoparticles},} {\protect\JournalTitle{J. Phys. Chem. C}}
  \textbf{111}, 9044--9048 (2007).

\bibitem{sveltobook}
O.~Svelto, \emph{Principles of lasers} (Springer, New York, NY, 2010), 5th ed.

\bibitem{Saad2021NL}
M.~S. Bin-Alam, J.~Baxter, K.~M. Awan, A.~Kiviniemi, Y.~Mamchur, A.~C. Lesina,
  K.~L. Tsakmakidis, M.~J. Huttunen, L.~Ramunno, and K.~Dolgaleva,
  \enquote{Hyperpolarizability of plasmonic meta-atoms in metasurfaces,}
  {\protect\JournalTitle{Nano Letters}} \textbf{21}, 51--59 (2021).

\end{thebibliography}

\end{document}